\documentclass[12pt]{article}
\usepackage{latexsym}
\usepackage{epsfig,amssymb,euscript,slashed}
\usepackage[linktocpage]{hyperref}
\usepackage{amsmath}
\usepackage{cite} 
\usepackage{array,calc,epsfig}
\usepackage{bbm}
\usepackage{fancybox}

\numberwithin{equation}{section} 
\usepackage[]{graphicx}
\usepackage{color}

\begin{document}
\font\cmss=cmss10 \font\cmsss=cmss10 at 7pt

\begin{flushright}{
\scriptsize QMUL-PH-18-31}
\end{flushright}
\hfill
\vspace{18pt}
\begin{center}
{\Large 
\textbf{Holographic correlators in AdS$_3$
}}

\end{center}

\vspace{8pt}
\begin{center}
{\textsl{Stefano Giusto$^{\,O^{++}, O^{--}}$, Rodolfo Russo$^{\,O^{+-}}$ and Congkao Wen$^{\,O^{+-}}$}}

\vspace{1cm}

\textit{\small ${}^{O^{++}}$ Dipartimento di Fisica ed Astronomia ``Galileo Galilei",  Universit\`a di Padova,\\Via Marzolo 8, 35131 Padova, Italy} \\  \vspace{6pt}

\textit{\small ${}^{O^{--}}$ I.N.F.N. Sezione di Padova,
Via Marzolo 8, 35131 Padova, Italy}\\
\vspace{6pt}

\textit{\small ${}^{O^{+-}}$ Centre for Research in String Theory, School of Physics and Astronomy\\
Queen Mary University of London,
Mile End Road, London, E1 4NS,
United Kingdom}\\
\vspace{6pt}

\end{center}

\vspace{12pt}

\begin{center}
\textbf{Abstract}
\end{center}

\vspace{4pt} {\small
  \noindent
  We derive the four-point correlators of scalar operators of dimension one in the supergravity limit of the D1D5 CFT holographically dual to string theory on AdS$_3\times S^3\times \mathcal{M}$, with $\mathcal{M}$ either $T^4$ or $K3$. We avoid the use of Witten diagrams but deduce our result from a limit of the heavy-heavy-light-light correlators computed in \cite{Galliani:2017jlg}, together with several consistency requirements of the OPE in the various channels. This result represents the first holographic correlators of single-trace operators computed in AdS$_3$.}

\vspace{1cm}

\thispagestyle{empty}

\vfill
\vskip 5.mm
\hrule width 5.cm
\vskip 2.mm
{
\noindent  {\scriptsize e-mails:  {\tt stefano.giusto@pd.infn.it, r.russo@qmul.ac.uk,  c.wen@qmul.ac.uk} }
}

\setcounter{footnote}{0}
\setcounter{page}{0}

\newpage


\section{Introduction}
Recent years have seen a remarkable progress in the computation and the analysis of conformal correlators at large 't Hooft coupling and large $N$, where the problem becomes tractable by using holographic techniques. For the paradigmatic example of four-dimensional $\mathcal{N}=4$ super Yang-Mills theory, dual to type IIB string theory on AdS$_5\times S^5$, we now know a general formula for the four-point correlators of half-BPS single-trace operators of arbitrary fixed dimension \cite{Rastelli:2016nze,Rastelli:2017udc}, and similar results are available for theories with an AdS$_7$ dual \cite{Rastelli:2017ymc}. Surprisingly, not a single holographic correlator of single-trace operators has ever been computed in AdS$_3$. One of the reasons is that the standard Witten diagram technique used to compute correlators in AdS$_{d+1}$ cannot be naively extrapolated to $d=2$: simply setting $d=2$ in the results for generic $d$ leads to divergent expressions for the diagrams computing the exchange of a massless vector and graviton~\cite{DHoker:1999mqo}, and an appropriate procedure to define the $d=2$ limit has not been formulated yet. Another issue is that the cubic couplings of type IIB supergravity on AdS$_3 \times S^3$ are known~\cite{Arutyunov:2000by}, but not the quartic ones which are also needed for deriving a full 4-point correlator by using Witten diagrams.

Here we will focus on 4-point correlators in AdS$_3$/CFT$_2$ of the type
\begin{equation}
\label{eq:corrg}
\langle O_1(z_1,\bar z_1) O_2(z_2,\bar z_2) O_3(z_3,\bar z_3) O_4(z_4,\bar z_4) \rangle = \frac{1}{|z_{12}|^{2 \Delta_1} |z_{34}|^{2\Delta_3}}\,\mathcal{G}(z,\bar z)\,,
\end{equation}
where we assume that the conformal dimensions satisfy $\Delta_1=\Delta_2$ and  $\Delta_3=\Delta_4$. As usual, global conformal invariance implies that the 4-point correlators can be expressed in terms of a function of the cross-ratio
\begin{equation}\label{eq:crossratio}
z\equiv \frac{z_{14} z_{23}}{z_{13} z_{24}}\,\,\Rightarrow \,\,1-z =\frac{z_{12} z_{34}}{z_{13} z_{24}}\,
\end{equation}
as done in~\eqref{eq:corrg}, so we will write our results in terms of the function ${\cal G}(z,\bar z)$.

Holographic correlators of the type~\eqref{eq:corrg} containing two light and two heavy operators, which we will dub HHLL correlators, have been studied from different points of view. Here the heavy states are described by multi-trace operators that have a conformal dimension proportional to the central charge $c$. The contribution to the HHLL correlators from the (large $c$) Virasoro block of the identity was derived in~\cite{Fitzpatrick:2015zha} within a CFT approach and in~\cite{Hijano:2015qja} from a bulk perspective. In order to go beyond this approximation and obtain a full HHLL CFT correlator it is necessary to specify the heavy operator more precisely. When the dimension of the heavy multi-trace operators is of order of the central charge, their gravitational backreaction effectively changes the background, replacing AdS$_3 \times S^3$ with a non-trivial smooth geometry which approximates AdS$_3 \times S^3$ only near the asymptotic boundary. The precise form of this regular solution encodes the choice of the heavy state. For half-BPS operators in the D1D5 CFT all such geometries are known \cite{Lunin:2001jy,Kanitscheider:2007wq}. The single-trace operators, on the other hand, are taken to have dimensions of order one in the large central charge limit: they thus represent perturbations of the background and are described by linear wave equations. Then, for the HHLL case, it is possible to calculate the four-point correlators of two single and two multi-trace operators bypassing the difficulties affecting Witten diagrams for $d=2$~\cite{Galliani:2016cai,Galliani:2017jlg,Bombini:2017sge}: they are extracted from a non-normalizable solution of the wave equation associated with the light operators in the geometry sourced by the heavy ones. This method effectively reduces the computation of the HHLL four-point function to that of a two-point function in a non-trivial background, and thus it does not require evaluating formally divergent exchange diagrams.

The question we address in this paper is how to reconstruct the correlators containing only light single-trace operators, which will be referred to as the LLLL correlators, from the HHLL correlators computed in \cite{Galliani:2017jlg,Bombini:2017sge}. This is possible because the multi-trace operators considered in \cite{Galliani:2017jlg,Bombini:2017sge} depend on a free parameter, denoted as $b$ in those references, which controls the number of their single-trace components and, hence, the dimension of the heavy operators can be made small by taking the parameter $b$ small. In this limit one thus naively expects that the HHLL correlators reduce to the LLLL ones. As we will see, this naive expectation is not quite correct. To understand where the problem is, we remark that in the HHLL correlators no single-trace operator is exchanged in the channels where a light and a heavy operator are fused together. Since this is true for any value of $b$, this feature survives the small $b$ limit. On the contrary the OPE between two single-trace operators does, in general, contain other single-trace operators. So, if we schematically denote by $O_L$ the single-trace operators appearing in the HHLL correlator and by $O'_L$ the single-trace operators obtained by taking the small $b$ limit of the original heavy operators, then the OPE of $O_L$ and $O'_L$ should, in general, contain the contribution of other single-trace operators, which is missed when we take the small $b$ limit of the HHLL correlator.  This implies that the result obtained from the HHLL correlator does not correctly describe the LLLL correlator in the limit in which $O_L$ and $O'_L$ are close. This problem, however, should not be relevant in the direct channel where the two $O_L$ operators are close: our fundamental assumption is that the small $b$ limit of the HHLL correlator correctly captures the contributions to the LLLL correlator due to single-trace operators
exchanged in this channel. We will show how the contributions from the other channels can be unambiguously fixed by various consistency requirements.

For example, when one of the $O_L$ operators is the same as one of the $O'_L$ operators, the symmetry of the correlator under the exchange of $O_L$ and $O'_L$ can be used to recover the correct OPE expansion in one of the crossed channels. More generally, we will require that the holographic correlators have the expected singularities associated to the exchange of protected single-trace operators in all OPE channels. 
This requirement determines the full correlator up to contact terms, which holographically correspond to Witten diagrams without internal propagators, and only affect the exchange of double-trace operators \cite{Heemskerk:2009pn}. In the examples we consider, the contact terms are fully determined by consistency with the flat space limit and by matching the contribution of some protected double-trace operator at weak and strong coupling. We apply this logic to the four-point function of half-BPS single-trace operators of dimension $(1/2,1/2)$ and find a single answer that passes all the consistency checks. We leave the generalization to correlators with higher dimensional operators for the future.

The paper is organised as follows. In section~\ref{sec:review} we characterise the half-BPS operators considered in this paper by using the free limit of the CFT description. Then we briefly review the results for the HHLL correlators in the supergravity limit that were derived in~\cite{Galliani:2017jlg,Bombini:2017sge}. In section~\ref{sec:reconstructing} we take the small $b$ limit and obtain, for the case of a particularly symmetric LLLL correlator, an expression that is valid in the direct channel limit $z_3 \to z_4$. Then we impose the consistency conditions mentioned above and obtain the result~\eqref{eq:corr1final} for the full LLLL correlator considered. Finally in section~\ref{sec:gen} we generalise this result to other correlators involving operators of the same dimension $(1/2,1/2)$, but with different R-symmetry charges. We check in a direct way that the results are consistent with the R-symmetry Ward identities and summarise all correlators we considered in a single equation~\eqref{eq:fr} where the dependence on the R-symmetry charges is factorised. In the final section \ref{sec:discussion} we present a brief discussion on how to use the Minkowskian inversion formula of~\cite{Caron-Huot:2017vep} to extract some large $N$ OPE data for the D1D5 CFT at strong coupling.

\section{Review of the known HHLL correlators}
\label{sec:review}

Exactly as it happens for ${\cal N}=4$ Super Yang-Mills, also the D1D5 Super CFT, which is dual to type IIB string theory on\footnote{Here ${\cal M}$ can be either $T^4$ or $K3$, as we will focus on a sector that is common to the two cases.} AdS$_3 \times S^3 \times {\cal M}$, has a free locus in its superconformal moduli space. On this locus the CFT is described by a collection free bosons and free fermions
\begin{equation}
  \{ \partial X^{A\dot{A}}_{(r)}(z),\, \psi^{\alpha \dot{A}}_{(r)} (z),\,
  \bar\partial X^{A\dot{A}}_{(r)}(\bar z),\, \tilde\psi^{\dot{\alpha} \dot{A}}_{(r)} (\bar z) \}~,
\end{equation}
where $r=1,\ldots, N = n_1 n_5$, $c=6N$, and $A\;,\dot{A}\;,\alpha\;,\dot{\alpha}$ are $SU(2)$ indices. To be precise, the free description is in terms of an orbifold CFT, where the orbifold group is the permutation group $S_N$ acting on the copy index $(r)$. The $SU(2)_L\times SU(2)_R$ acting respectively on $\alpha$ and $\dot\alpha$ are part of the (affine) R-symmetry of the SCFT and will play an important role in our analysis. As usual we can characterise the (protected) chiral primary operators in terms of this free field description and in this work we focus on the following particularly simple operator
\begin{equation}
  \label{eq:Osing}
  O^{\alpha\dot\alpha} = \sum_{r=1}^N \frac{-i \epsilon_{\dot{A} \dot{B}}}{\sqrt{2 N}} \psi_{(r)}^{\alpha \dot{A}}\, \tilde{\psi}_{(r)}^{\dot\alpha \dot{B}}~.
\end{equation}
This operator is in the untwisted sector of the orbifold CFT and is manifestly symmetric under $S_N$; it is straightforward to see that it is chiral primary operator, since its conformal weights are $(h,\bar{h})=(1/2,1/2)$ and its charges under the generators $(J^3,\tilde{J}^3)$ in $SU(2)_L\times SU(2)_R$ are  $(j,\bar{j})=(\pm 1/2,\pm 1/2)$ depending on the values of $\alpha$ and $\dot{\alpha}$. On the bulk side of the duality~\eqref{eq:Osing} corresponds to a supergravity fluctuation of AdS$_3 \times S^3$, so we will refer to this type of operators as single particle or equivalently single trace in analogy with the nomenclature used in the case of ${\cal N}=4$ Super Yang-Mills.

An important property of 2d CFTs with extended supersymmetry is the possibility to perform a spectral flow transformation. Here we follow the conventions of~\cite{Giusto:2013bda}, then the minimal spectral flow, connecting the NSNS and the RR sector of theory, maps the NSNS $SL(2,\mathbb{C})$ invariant vacuum into the RR ground state with $(j,\bar{j})=(N/2,N/2)$ and of course $(h,\bar{h})=(c/24,c/24)$. This RR state is heavy according to the characterisation discussed in the introduction and is described in the dual bulk picture by a supersymmetric solution of type IIB supergravity~\cite{Balasubramanian:2000rt,Maldacena:2000dr,Lunin:2001fv}. Here we are interested in more general RR ground states that are obtained by taking the spectral flow of multiparticle antichiral primary operators. Before taking the spectral flow, these states have $(j,\bar{j})=(-h,-\bar{h}$) and are obtained just by inserting many copies of the same single-particle antichiral primary operator at the same point. In particular we are interested in the multiparticle state obtained by multiplying the operator~\eqref{eq:Osing} $N_b$ times
\begin{equation}
  \label{eq:Omult}
  O^{\rm NSNS}_{b} = \left(O^{++}\right)^{N_b}\;,~~~  \bar{O}^{\rm NSNS}_{b} = \left(O^{--}\right)^{N_b}\;.
\end{equation}
After spectral flowing this state to the RR sector one obtains a heavy state ($O^{\rm NSNS}_{b}\to \bar{O}_H$ and $\bar{O}^{\rm NSNS}_{b}\to {O}_H$, according to the conventions of~\cite{Galliani:2017jlg}) that is described by another supersymmetric type IIB solution\footnote{To be precise the supergravity solution is dual to a coherent state with a weighted sum over all possible numbers of single particle constituents which, in the large $N$ limit, is sharply peaked~\cite{Skenderis:2006ah}.}, see for instance Eqs.~(3.1) and~(B.1) of~\cite{Galliani:2017jlg}. The only feature of this solution we need to recall here is that it depends on a continuous parameter $b$ which is related to the number of single particle constituents in the heavy state (when considered before the spectral flow).

In the supergravity description the operator~\eqref{eq:Osing} corresponds to a fluctuation of the background encoded in a scalar and a 3-form of the 6d supergravity obtained from the standard KK reduction of type IIB on ${\cal M}$~\cite{Galliani:2017jlg}. By studying the equations of motion of this perturbation in the background, it is possible to extract the HHLL correlators $\langle O_H \bar{O}_H O^{\alpha\dot\alpha} O^{\beta\dot\beta}\rangle$, see\footnote{Of course these results are reliable in the supergravity regime where one takes $N$ to be large and an appropriate strong coupling limit.} Eqs.~(3.58) and~(F.3)~\cite{Galliani:2017jlg}. For convenience we report those results here after spectral flowing back to the NSNS sector where the heavy operators become the corresponding multiparticle states: the correlator $\langle \bar{O}^{\rm NSNS}_b {O}^{\rm NSNS}_b O^{++} O^{--}\rangle$ reads
\begin{equation}
  \label{eq:3.58}
  \mathcal{G}_{HHLL}(z,\bar z) = \left[1+\frac{b^2}{a_0^2}\,\left(\frac{|z|^2}{\pi}\,\hat D_{2211}-\frac{1}{2}+\frac{N}{2}\,|1-z|^2\right)\right]+ {\cal O}(b^4)\,,
\end{equation}
while for the correlator $\langle \bar{O}^{\rm NSNS}_b {O}^{\rm NSNS}_b O^{+-} O^{-+}\rangle$ we have
\begin{equation}
  \label{eq:F.3}
  \mathcal{G}_{HHLL}(z,\bar z) = \left[1+\frac{b^2}{a_0^2}\,\left(\frac{z}{\pi}\,\hat D_{2211}-\frac{1}{2}\right)\right]+ {\cal O}(b^4)\,,
\end{equation}
where $b^2/(2 a_0^2) =N_b/N$ and the definition of the $D$-functions is summarised in appendix~\ref{sec:apphatD}. In momentum space it is also possible to write an explicit expression for these correlators that is exact in $b$~\cite{Bombini:2017sge}, but we will not need it here since we will focus on the small $b$ limit.

\section{Reconstructing the LLLL correlators}
\label{sec:reconstructing}
The general picture that has emerged from the recent developments in holographic correlators \cite{Rastelli:2016nze,Alday:2016njk,Caron-Huot:2017vep,Alday:2017gde,Rastelli:2017udc,Aprile:2017bgs,Alday:2017xua,Alday:2017vkk,Caron-Huot:2018kta} is that the contributions of the single-trace operators\footnote{More precisely the contributions of the single-trace operators that are degenerate with double-trace operators are not needed, and hence all the information on holographic correlators is obtained, in the supergravity limit, from a finite number of terms.} exchanged in the various channels, together with the constraints coming from supersymmetry, determine the correlators uniquely. We will adopt a similar approach here and reconstruct the correlators with four single-trace operators from the information extracted from the HHLL correlators reviewed above, supplemented by some basic consistency requirements of the OPE in the various channels. 

We start from the analysis of the small $b$ limit of the HHLL correlator \eqref{eq:3.58} which should capture the contributions of the single-trace operators exchanged in the $s$-channel $z\to 1$ (i.e. $z_1\to z_2$) for the correlator
\begin{equation}
\label{eq:corr1}
\langle O^{--}(z_1,\bar z_1) O^{++}(z_2,\bar z_2) O^{++}(z_3,\bar z_3) O^{--}(z_4,\bar z_4) \rangle = \frac{1}{|z_{12} z_{34}|^2}\,\mathcal{G}(z,\bar z)\,.
\end{equation}
Naively one would expect to recover this LLLL correlator by setting $N_b=N b^2/(2 a_0^2)=1$ in the HHLL result \eqref{eq:3.58}. It is however clear that this cannot lead to the correct answer: for once, the $N_b=1$ limit of \eqref{eq:3.58} is not symmetric\footnote{Only the leading term in $N$, given by $1+|1-z|^2$, has the expected symmetry.} under the exchange of $z_2$ and $z_3$, as one can verify using the property \eqref{eq:dhat3} of the $\hat D$-functions, while the LLLL correlator in \eqref{eq:corr1} obviously is.  In hindsight, the failure of this naive expectation is not surprising: the HHLL result \eqref{eq:3.58} has been derived by first taking  $N_b$ to scale like $N$ in the large $N$ limit (so that $O_b^{NSNS}$ can be treated as a heavy operator and be represented by a geometry), and then by sending the ratio $N_b/N$ to zero; the LLLL correlator, on the other hand, should be computed by setting $N_b=1$ from the start, and then sending $N$ to infinity. There is a priori no reason that the these two different limits agree, and indeed they do not. The question is if we can still learn something on the LLLL correlator from \eqref{eq:3.58}. Our fundamental assumption will be that the small $b$ limit of the HHLL correlator correctly captures only the contributions to the LLLL correlator associated with the exchange of single-trace operators in the $s$-channel. If we denote by $\mathcal{G}_s(z,\bar z)$ this contribution, we then deduce from \eqref{eq:3.58} that
\begin{equation}
\mathcal{G}_s(z,\bar z) = 1 +\frac{1}{N}\,\left[\frac{2}{\pi} |z|^2 |1-z|^2 \hat D_{1122} -1\right]\,.
\end{equation}
Note that we have not included in the expression above the term $|1-z|^2$, which comes from the exchange of the identity in the $u$-channel, and, for later convenience, we have expressed $\hat D_{2211}$ in terms of $\hat D_{1122}$, using the identity $\hat D_{2211}=|1-z|^2 \hat D_{1122}$ which follows from \eqref{eq:Id12}.

A consistency check on the above form for $\mathcal{G}_s$ is obtained by looking at the light-cone limit $\bar z\to 1$, where one expects that the only exchanged states are the affine descendants of the identity (i.e.~the descendants generated by the Virasoro and the R-symmetry generators), since these are the only protected states with $\bar h=0$. One can indeed check that in this limit
\begin{equation}\label{eq:idblock}
\mathcal{G}_s \stackrel{\bar z\to 1}{\longrightarrow} 1 -\frac{1}{N}\,\left(1+\frac{z}{1-z} \log z\right)\,,
\end{equation}
which is the identity affine block\footnote{This can be seen as follows: The identity Virasoro block at order $1/N$ is equivalent to the global block of the stress tensor; for external states of dimension $h_L=1/2$ this is 
\begin{equation}\nonumber
\mathcal{V}_V= 1+ \frac{1}{12 N} (1-z)^2 \,{}_2F_1(2,2,4;1-z)=1-\frac{1}{N} \left[ 1+\frac{1+z}{2(1-z)}\log z\right]\,;
\end{equation}
the $U(1)$-affine block for external states of charge $q_L=1/2$ is (see for instance \cite{DiFrancesco:1997nk})
\begin{equation}\nonumber
\mathcal{V}_A = z^\frac{1}{2N}=1+\frac{1}{2 N}\log z+\mathcal{O}(N^{-2})\,.
\end{equation}
The identity affine block is then
\begin{equation}\nonumber
\mathcal{V} = \mathcal{V}_V \mathcal{V}_A = 1-\frac{1}{N}\,\left(1+\frac{z}{1-z} \log z\right)+\mathcal{O}(N^{-2})\,.
\end{equation}} expanded up to order $1/N$. 

In the full correlator \eqref{eq:corr1}, single-trace operators are also exchanged in the $u$-channel $z\to \infty$ (or $z_1\to z_3$). This $u$-channel contribution is easily determined by imposing the symmetry under the exchange of $z_2$ and $z_3$:
\begin{equation}
\frac{1}{|z_{12} z_{34}|^2}\,\mathcal{G}(z,\bar z) = \frac{1}{|z_{13} z_{24}|^2}\,\mathcal{G}(z',\bar z')\quad \mathrm{with}\quad z'\equiv -\frac{z_{14} z_{23}}{z_{12} z_{34}}= \frac{z}{z-1}\,,
\end{equation}
which implies
\begin{equation}\label{eq:z2z3sym}
\mathcal{G}(z,\bar z)=|1-z|^2 \,\mathcal{G}\left(\frac{z}{z-1},\frac{\bar z}{\bar z-1} \right)\,.
\end{equation}
An obvious way to obtain a correlator with the correct symmetry is to add to $\mathcal{G}_s$ the term
\begin{equation}
\mathcal{G}_u(z,\bar z)=|1-z|^2 \,\mathcal{G}_s\left(\frac{z}{z-1},\frac{\bar z}{\bar z-1} \right) = |1-z|^2 +\frac{|1-z|^2}{N}\,\left[\frac{2}{\pi} |z|^2 \hat D_{1212} -1\right]\,.
\end{equation}
In deriving the above equation we have used the transformation property of the $\hat D$-functions explained in \eqref{eq:dhat3}.

Since in the $t$-channel $z\to 0$ (or $z_1\to z_4$) there is no exchange of single-trace operators with dimensions less than $(h ,\bar h)=(1,1)$, at which the first double-trace operator appears, the sum of $\mathcal{G}_s$ and $\mathcal{G}_u$ should correctly include all the single-trace exchanges. This, however, does not completely identify the correlator, as one still has the freedom to add terms that only affect the contributions of the double-trace operators. From the bulk point of view, such terms originate from Witten diagrams associated to quartic contact interactions, and have first been studied in \cite{Heemskerk:2009pn}. They are in principle determined by supersymmetry, which completely fixes the supergravity action at the two-derivative level. We follow here a simpler route, and resolve this ambiguity by requiring consistency with the flat space limit and with the $t$-channel OPE. The constraint coming from the flat space limit is easier to formulate in Mellin space \cite{Mack:2009mi, Penedones:2010ue}: the allowed contact interactions should be functions of the Mellin variables scaling at most linearly when all the variables are taken to infinity. Moreover, the requirement that the contact interactions do not introduce single-trace exchanges implies that their Mellin transform have no poles. One can show that these constraints, together with the symmetry under the exchange of the $s$ and $u$-channels, leave only two possible contact terms. In Mellin space, they are proportional to $t$ and a constant. As we show in appendix \ref{app:Mellin}, they correspond to the following two contributions\footnote{The fact that $\mathcal{G}_\mathrm{cont.}$ in \eqref{eq:Gcont} satisfies the symmetry \eqref{eq:z2z3sym} follows from \eqref{eq:dhat3}.}, when re-expressed in the space-time coordinates:
\begin{equation}
  \label{eq:Gcont}
\mathcal{G}_\mathrm{cont.}(z,\bar z)=\frac{2}{\pi\,N}\,|1-z|^2 \left(c_1 \hat D_{1111} +  c_2 |z|^2 \hat D_{2112}\right)\,.
\end{equation}
We can fix\footnote{We thank Agnese Bissi for suggesting this possibility to us.} the coefficients $c_1$ and $c_2$ by looking at the $z\to 0$ limit of the correlator: the lowest dimension operator exchanged in this channel is the double-trace $:O^{++} O^{++}:$, which results from the fusion of two chiral primaries of the same chirality and, hence, it must have a vanishing anomalous dimension and its three-point coupling with the external operators should be the same as in the free theory. The vanishing of the anomalous dimension requires that the correlator does not contain terms of the type $\log|z|^2$ for $z\to 0$: it is immediate to see that this condition imposes $c_1=0$. One should also require that the coefficient of the term of order $(z \bar z)^0$, which encodes the square of the three-point function $\langle O^{--} O^{--} :O^{++} O^{++}:\rangle$, equals the free theory result. The free correlator is given by
\begin{equation}\label{eq:fr0pp}
\mathcal{G}_\mathrm{free}(z,\bar z)=1+|1-z|^2 + \frac{1}{N}\,\frac{|z|^2-1-|1-z|^2}{2}= 2-\frac{1}{N}+\mathcal{O}(z)+\mathcal{O}(\bar z)\,,
\end{equation}
while expanding the correlator at strong coupling (with $c_1=0$) one finds
\begin{equation}
\mathcal{G}_s(z,\bar z)+\mathcal{G}_u(z,\bar z)+\mathcal{G}_\mathrm{cont.}(z,\bar z) = 2-\frac{2-c_2}{N}+\mathcal{O}(z)+\mathcal{O}(\bar z)\,,
\end{equation}
and thus this determines $c_2=1$. 

Our final result for the correlator \eqref{eq:corr1} is then
\begin{equation}\label{eq:corr1final}
\mathcal{G}(z,\bar z) = \left(1-\frac{1}{N}\right)(1+|1-z|^2)+\frac{2}{\pi\,N}\,|z|^2 |1-z|^2 (\hat D_{1122}+\hat D_{1212}+\hat D_{2112}) \,.
\end{equation}

\section{The full R-symmetry multiplet}
\label{sec:gen}

One can generalize the result of the previous section by considering correlators of general operators $O^{\alpha \dot \alpha}$ in the same R-symmetry multiplet. We can introduce the general notation
\begin{equation}
\langle O^{\alpha_1 \dot \alpha_1}(z_1,\bar z_1) O^{\alpha_2 \dot \alpha_2}(z_2,\bar z_2) O^{\alpha_3 \dot \alpha_3}(z_3,\bar z_3) O^{\alpha_4 \dot \alpha_4}(z_4,\bar z_4) \rangle = \frac{1}{|z_{12} z_{34}|^2}\,\mathcal{G}^{(\alpha_1\dot\alpha_1) (\alpha_2 \dot\alpha_2)}_{(\alpha_3\dot\alpha_3) (\alpha_4 \dot\alpha_4)}(z,\bar z)\,,
\end{equation}
so that the correlator in \eqref{eq:corr1} (or \eqref{eq:corr1final}) will be renamed $\mathcal{G}^{(--) (++)}_{(++) (--)}$. Essentially the only other correlator to compute is $\mathcal{G}^{(--) (++)}_{(+-) (-+)}$: we could deduce this correlator from its HHLL ``parent'' \eqref{eq:F.3}, using arguments similar to the ones outlined in the previous section, or by using the Ward identity that relates correlators within the same R-symmetry multiplet. We will follow this second route in this section.

A straightforward way to derive the Ward identity is to write one of the $O^{--}$ operators in the correlator \eqref{eq:corr1} as $O^{--}=[J^-_0,O^{+-}]$ and to move the current $J^-_0$ on the other operators by the usual argument of deforming the integration contour. Using $[J^-_0,O^{--}]=0$ and $[J^-_0,O^{++}]=-O^{-+}$ (where the minus sign is needed to have $O^{-+}= (O^{+-})^\dagger$), one immediately finds
\begin{equation}\label{eq:WI}
\begin{aligned}
\mathcal{G}^{(--) (++)}_{(++) (--)}(z,\bar z)&=\mathcal{G}^{(--) (++)}_{(-+) (+-)}(z,\bar z)+\mathcal{G}^{(--) (-+)}_{(++) (+-)}(z,\bar z)\\
&=\mathcal{G}^{(--) (++)}_{(+-) (-+)}\left(\frac{1}{z},\frac{1}{\bar z}\right)+|1-z|^2\, \mathcal{G}^{(--) (++)}_{(+-) (-+)}\left(\frac{z-1}{z},\frac{\bar z-1}{\bar z}\right)\,,
\end{aligned}
\end{equation}
where in the second step we have expressed the r.h.s. in terms of the single correlator $\mathcal{G}^{(--) (++)}_{(+-) (-+)}$ by exchanging the positions of the operators. One can check, using the transformation properties of the $\hat D$-functions in \eqref{eq:Id12} and \eqref{eq:dhat3}, that a solution of the Ward identity \eqref{eq:WI} is
\begin{equation} \label{eq:correlator2}
\mathcal{G}^{(--) (++)}_{(+-) (-+)}(z,\bar z)=1-\frac{1}{N}+\frac{2}{\pi\,N}\,z\, |1-z|^2 (\hat D_{1122}+\hat D_{1212}+\hat D_{2112}) \,.
\end{equation}
The above solution passes several consistency checks: it reduces to the appropriate identity affine blocks in the light-cone limits $\bar z\to 1$ and $z\to 1$  (note that due to the opposite R-charges of the external operators in the left and right sectors, the contribution from the R-symmetry block is $z^\frac{1}{2N}$ for $\bar z\to 1$ and $\bar z^{-\frac{1}{2N}}$ for $z\to 1$), and the contributions from the exchange of protected operators in the various channels agree with the free correlator 
\begin{equation}\label{eq:fr0pm}
\mathcal{G}^{(--) (++)}_{(+-) (-+)\,\mathrm{free}}(z,\bar z)=1+\frac{1}{2 N}\left(\frac{|1-z|^2}{\bar z}+\frac{1-\bar z}{\bar z}-(1-z)\right)\,.
\end{equation}
Note that for this correlator the first non-protected operators appear at dimension $(h,\bar h)=(1,1)$ in the $s$-channel, but only at dimension $h+\bar h\ge 3$ in the $t$ and $u$-channels, because multi-trace operators like $:O^{++} O^{+-}:$ or $:O^{++} O^{-+}:$ preserve supersymmetry in the left or in the right sector and are thus protected. We believe that there are no contact terms with correct symmetries that could be added to $\mathcal{G}^{(--) (++)}_{(+-) (-+)}$ without spoiling the Ward identity \eqref{eq:WI} or modifying the contributions from the exchange of protected operators. 

Analogue to the case of $\mathcal{N}=4$ SYM, to keep track of the R-symmetry it is convenient to introduce two-dimensional vectors and to define 
\begin{align}
O_i = A^{i}_{\alpha} \bar{A}^{i}_{\dot \alpha} O^{\alpha \dot{\alpha}} (z_i, \bar{z}_i),
\end{align}
where, for each value of $i$, the vectors $A^{i}_{\alpha}$ are given by either
\begin{align}
A_{+} = \begin{pmatrix} 
1 \\
0 
\end{pmatrix}\quad \mathrm{or} \quad 
A_{-} = \begin{pmatrix} 
0 \\
1 
\end{pmatrix},
\end{align}
and the same for  $\bar{A}^i_{\dot \alpha}$.  With this set up, the general four-point correlator of the R-symmetry multiplet is now defined as
\begin{align}
\langle O_1 O_2 O_3 O_4 \rangle = {|A^1\cdot A^2 A^3\cdot A^4|^2  \over |z_{12} \, z_{34}|^2} \, \mathcal{G}(\alpha_c, \bar{\alpha}_c, z, \bar{z}),
\end{align}
where $\alpha_c$ is the cross ratio
\begin{align}
\alpha_c = {A^1\cdot A^3 A^2\cdot A^4 \over A^1\cdot A^4 A^2\cdot A^3},
\end{align}
and similarly for $\bar{\alpha}_c$. The dot in the above formulas denotes the contraction with $\epsilon_{\alpha \beta}$ or $\epsilon_{\dot \alpha \dot \beta}$.  With the help of these R-symmetry variables, we find that the free correlators can now be expressed as
\begin{align}\label{eq:fr0a}
 \mathcal{G}_{\rm free}(\alpha_c, \bar{\alpha}_c, z, \bar{z}) &= 
1 +   {|1 - z|^2 \over |1-{\alpha}_c |^2 }\left(|{\alpha}_c|^2 + { 1\over |z|^{2} } \right)  + \cr
&+\, 
{1\over 2N }\left( {1 - z \over z(1 - \alpha_c)} + {\alpha_c (1 - z)  \over 1 - \alpha_c} 
- {\alpha_c  |1 - z|^2 \over  \bar{z} |1 - \alpha_c|^2 } + c.c. \right)
\end{align}
where $c.c.$ represents the conjugate terms with $\alpha_c \leftrightarrow \bar{\alpha}_c, z \leftrightarrow \bar{z}$. Whereas for the $AdS$ correlator $\mathcal{G}(\alpha_c, \bar{\alpha}_c, z, \bar{z})$, the main interest of the paper, we find that the result is given by
\begin{align}
  \label{eq:fr}
  \mathcal{G}(\alpha_c, \bar{\alpha}_c, z, \bar{z}) =\mathcal{G}_0 +  {1\over N}{ |1 - {\alpha}_c \, z|^2 \over |1-{\alpha}_c |^2 }  \left[ {2 \over \pi}|1-z|^2 
\left( \hat{D}_{1122} +  \hat{D}_{1212}+  \hat{D}_{2112}  \right)  \right],
\end{align}
where $\mathcal{G}_0$ contains only rational functions of $z, \bar{z}$ and is given by
\begin{align}
  \label{eq:fr0}
  \mathcal{G}_0 =\left(1 - {1 \over N} \right) \left[ 1 +   {|1 - z|^2 \over |1-{\alpha}_c |^2 }\left(|{\alpha}_c|^2 + { 1\over |z|^{2} } \right) \right].
 \end{align}
 Note the leading-$N$ term comes from $\mathcal{G}_{\rm free}$ in (\ref{eq:fr0a}). It is straightforward to check that $\mathcal{G}(\alpha_c, \bar{\alpha}_c, z, \bar{z})$ and $ \mathcal{G}_{\rm free}(\alpha_c, \bar{\alpha}_c, z, \bar{z}) $ have all the correct symmetries and that, in the limits of $\{ {\alpha}_c \rightarrow -\infty, \bar{\alpha}_c \rightarrow -\infty \}$ and $\{ {\alpha}_c \rightarrow -\infty, \bar{\alpha}_c \rightarrow 0 \}$, \eqref{eq:fr} reduces to (\ref{eq:corr1final}) and (\ref{eq:correlator2}) while ~\eqref{eq:fr0a} reduces to~\eqref{eq:fr0pp} and~\eqref{eq:fr0pm}. Finally, we comment that both $ \mathcal{G}_{\rm free}(\alpha_c, \bar{\alpha}_c, z, \bar{z}) $ and $\mathcal{G}(\alpha_c, \bar{\alpha}_c, z, \bar{z})$ satisfy 
\begin{align}
\partial_{\bar z} \left( \mathcal{G}_{\rm free}(\alpha_c, \bar{\alpha}_c, z, \bar{z})  \big{|}_{\bar{\alpha} \rightarrow 1/\bar{z}} \right) =0, \quad \quad \partial_{\bar z} \left( \mathcal{G}(\alpha_c, \bar{\alpha}_c, z, \bar{z})  \big{|}_{\bar{\alpha} \rightarrow 1/\bar{z}} \right) =0
 \end{align}
 which takes exactly the same form as  the superconformal Ward identity in the case of $\mathcal{N}=4$ SYM~\cite{Eden:2000bk, Nirschl:2004pa}. 

\section{A first look at the anomalous dimensions}
\label{sec:discussion}

The main result of this note is summarised in eqs.~\eqref{eq:fr}, \eqref{eq:fr0}, which give the holographic correlators of the scalar operators of dimension $(1/2,1/2)$ defined, at the free point of the CFT, in \eqref{eq:Osing}. We hope that this result may pave the way for a systematic construction of holographic correlators in AdS$_3$.

In the much better studied AdS$_5$ case, the knowledge of holographic correlators has lead to uncover a very rich structure, in particular for what concerns the spectrum of double-trace operators of $\mathcal{N}=4$ SYM at strong coupling (see for example \cite{Alday:2017gde,Aprile:2017xsp,Aprile:2018efk,Caron-Huot:2018kta}). An analogous information should be encoded also in AdS$_3$ correlators. Considering for example the correlator in \eqref{eq:corr1}, the $s$-channel OPE contains a series of double-trace operators of the form 
\begin{equation}
O_{m, \bar m} = \,:\!O^{--} \partial^m \bar \partial^{\bar m} O^{++}\!:\quad \mathrm{with} \quad m,\bar m=0,1,\ldots\,,
\end{equation} 
whose conformal dimensions $(h_{m,\bar m},\bar h_{m,\bar m})$ receive quantum corrections at order $1/N$:
\begin{equation}
h_{m,\bar m} = 1+m + \frac{\gamma_{m,\bar m}}{N}\,,\quad {\bar h}_{m,\bar m} = 1+{\bar m} + \frac{\gamma_{m,\bar m}}{N}\,.
\end{equation}
As usual the anomalous dimensions $\gamma_{m,\bar m}$ can be extracted from the terms containing $\log|1-z|^2$ in the $z\to1$ expansion of the correlator. Actually things are more involved: in the CFT there are other operators with the same bare dimension as $O_{m, \bar m}$, and all these operators could mix away from the free orbifold point. The knowledge of the correlators derived in this note is not enough to resolve this mixing problem. The problem simplifies considerably if we limit ourselves to compute the anomalous dimensions averaged over all the operators with the same bare dimension, which in the following we will denote as $\langle \gamma_{m,\bar m}\rangle$. In general, this computation could be performed using two equivalent approaches \cite{Alday:2017vkk}: the large spin perturbation theory of \cite{Alday:2016njk,Alday:2017gde,Alday:2017xua} or the Lorentzian inversion formula of \cite{Caron-Huot:2017vep,Caron-Huot:2018kta}. The power of these methods is that they allow to deduce the averaged OPE data in one channel solely from the singularities of the correlator in the crossed channels which at large $N$ are determined by the exchange of the protected single trace operators. Thus, to compute the anomalous dimensions of the double-trace operators exchanged in the $s$-channel of the correlator in \eqref{eq:corr1}, one has to consider only the single-trace contributions in the $u$-channel (since in this correlator no single-particle operator is exchanged in the $t$-channel). As explained in section~\ref{sec:reconstructing}, this protected part is just the affine block of the identity. The derivation of the averaged anomalous dimensions in a two-dimensional correlator where the identity and the graviton are the only exchanged single-particle states has already been performed in \cite{Kraus:2018zrn} using the Lorentzian inversion method. The only difference with our correlator is that we also have the contribution of the R-symmetry currents: it is easy to see how this new contribution modifies the relevant integrand in \cite{Kraus:2018zrn} and so we can obtain the result relevant for our correlator~\eqref{eq:corr1} from that of \cite{Kraus:2018zrn}. We find
\begin{equation}\label{eq:avgamma}
\langle \gamma_{m,\bar m} \rangle=-(n^2+n+1)\quad \mathrm{with}\quad n = \mathrm{min}(m,\bar m)\,.
\end{equation}
We verified that these agree with the anomalous dimensions derived by directly expanding the correlator \eqref{eq:corr1final} for $z\to 1$ when the spin of the double-trace operator is not too small: $\ell=|m-\bar m|>2$. It is known \cite{Caron-Huot:2017vep,Simmons-Duffin:2017nub} that in the large $N$-expansion the Lorentzian inversion formula can fail to work for $\ell=0,1,2$, and indeed we find that the simple formula \eqref{eq:avgamma} fails to reproduce the averaged anomalous dimensions for these values of $\ell$. For example from the $z\to 1$ expansion of the Euclidean correlator we find $\langle \gamma_{0,0} \rangle=\langle \gamma_{1,0} \rangle=\langle \gamma_{0,1} \rangle=-5/6$, $\langle \gamma_{2,0} \rangle=\langle \gamma_{0,2} \rangle=-14/15$.

As we mentioned above, the solution of the mixing problem and the full derivation of the double-trace spectrum requires the knowledge of a more general set of correlators. We hope to report progress in this direction in the future.

\section*{Acknowledgements}

We would like to thank Agnese Bissi and Leonardo Rastelli for discussions and correspondence, and Luis F. Alday for several helpful comments on a preliminary verion of this paper. This work was partially supported in part by the Science and Technology Facilities Council (STFC) Consolidated Grant ST/L000415/1 {\it String theory, gauge theory \& duality}. C.W. is supported by a Royal Society University Research Fellowship No. UF160350

\appendix
\section{The $\hat D$-functions}
\label{sec:apphatD}
The contact Witten diagram in AdS$_{d+1}$ with external operators of dimension $\Delta_i$, usually denoted as the $D$-function, is given by
\begin{equation}
\begin{aligned}
D_{\Delta_1\Delta_2\Delta_3\Delta_4}(\vec{z}_1,\vec{z}_2,\vec{z}_3,\vec{z}_4)&=\int d^{d=1}{w}\, \sqrt{g} \,\prod_{i=1}^4 K_{\Delta_i}(w;\vec{z}_i)\\
&=\Gamma\!\left(\frac{\hat \Delta-d}{2}\right)\frac{\pi^{d/2}}{2}\int_0^\infty \prod_i\left[dt_i \frac{t_i^{\Delta_i-1}}{\Gamma(\Delta_i)}\right]\,e^{\sum_{i,j=1}^4 z_{ij}^2\,\frac{t_i t_j}{2}}\,,
\end{aligned}
\end{equation}
where $\hat \Delta = \sum_i \Delta_i$, $z_{ij}^2 = (\vec{z}_i - \vec{z}_j)^2$, $g$ is the determinant of the AdS$_{d+1}$ metric in Euclidean Poincar\'e coordinates $w\equiv (w_0, \vec{w})$
\begin{equation}
ds^2 = \frac{dw_0^2 + \sum_{i=1}^d dw_i^2}{w_0^2}\,,
\end{equation}
and  $K_{\Delta}(w;\vec{z})$ is the bulk-to-boundary propagator for a scalar field of conformal dimension $\Delta$:
\begin{equation}
K_\Delta(w,\vec{z})=\left[ \frac{w_0}{w_0^2+(\vec{w}-\vec{z})^2}\right]^\Delta\,,
\end{equation}
with $\vec{z}$, $\vec{w}$ points on the $d$-dimensional boundary. One can define $\bar D$-functions, which are independent of the dimension $d$ and depend on the cross-ratios $z$ and $\bar z$
\begin{equation}
(1-z)(1-\bar z)=\frac{z_{12}^2 z_{34}^2}{z_{13}^2 z_{24}^2} \,,\quad z \bar z\ =\frac{z_{14}^2 z_{23}^2}{z_{13}^2 z_{24}^2} \,,
\end{equation}
as
\begin{equation}
\bar D_{\Delta_1\Delta_2\Delta_3\Delta_4}(z,\bar z)=\frac{2 \prod_{i=1}^4 \Gamma(\Delta_i)}{\pi^{d/2}\Gamma\!\left(\frac{\hat \Delta-d}{2}\right)}\frac{|z_{13}|^{\hat \Delta-2\Delta_4}\,|z_{24}|^{2 \Delta_2}}{|z_{14}|^{\hat \Delta - 2\Delta_1 - 2\Delta_4} |z_{34}|^{\hat \Delta -2 \Delta_3-2\Delta_4}} D_{\Delta_1\Delta_2\Delta_3\Delta_4}(\vec{z}_1,\vec{z}_2,\vec{z}_3,\vec{z}_4)\,.
\end{equation}
The $\hat D$-functions which we use in the bulk of the article are instead defined in terms of the $D$-functions with $d=2$ as
\begin{equation}
\hat D_{\Delta_1\Delta_2\Delta_3\Delta_4}(z,\bar z)=\lim_{z_2\to \infty}|z_2|^{2\Delta_2} D_{\Delta_1\Delta_2\Delta_3\Delta_4}(z_1=0,z_2,z_3=1,z_4=z)\,,
\end{equation}
where it is understood that we parametrize a 2-dimensional point $\vec{z}_i$ by the complex number $z_i$. The relation between $\hat D$ and $\bar D$-functions is thus
\begin{equation}
\hat D_{\Delta_1\Delta_2\Delta_3\Delta_4}(z,\bar z) = \frac{\pi\,\Gamma\!\left(\frac{\hat \Delta-2}{2}\right)}{2 \prod_{i=1}^4 \Gamma(\Delta_i)}\,|z|^{\hat \Delta - 2\Delta_1 - 2\Delta_4} |1-z|^{\hat \Delta -2 \Delta_3-2\Delta_4}\bar D_{\Delta_1\Delta_2\Delta_3\Delta_4}(z,\bar z)\,.
\end{equation}
The $\hat D$-functions relevant for this article can be reconstructed from
\begin{equation}
\hat D_{1111}(z,\bar z)=\frac{2 \pi i}{z-\bar z} D(z,\bar z)\,,
\end{equation}
and
\begin{equation}
\hat D_{1122}(z,\bar z) = -\frac{2\pi i}{(z-\bar z)^2}\,\left[ \frac{z+\bar z}{z-\bar z} D(z,\bar z)+\frac{\log|1-z|^2}{2i}+ \frac{z+\bar z-2 z \bar z}{4i \,|1-z|^2}\log|z|^2\right]\,,
\end{equation}
where $D(z,\bar z)$ is the Bloch-Wigner dilogarithm
\begin{equation}
D(z,\bar z)= \frac{1}{2 i} \left[ \mathrm{Li}_2(z) - \mathrm{Li}_2(\bar z) +\frac{1}{2} \log|z|^2 \log \left(\frac{1-z}{1-\bar z}\right)\right]\,.
\end{equation}
The $\hat D$-functions have simple transformation properties under exchange of the various points $z_i$. The identities used in the article are
 \begin{align} \label{eq:Id12}
\hat{D}_{\Delta_2 \Delta_1 \Delta_3 \Delta_4}\left({1\over z}, {1\over \bar{z}} \right) &= |z|^{2 \Delta_4} 
\hat{D}_{\Delta_1 \Delta_2 \Delta_3 \Delta_4}(z, {\bar{z}} )\, , \cr 
\hat{D}_{\Delta_3 \Delta_2 \Delta_1 \Delta_4}({1- z}, {1- \bar{z}} ) &= 
\hat{D}_{\Delta_1 \Delta_2 \Delta_3 \Delta_4}({z}, {\bar{z}} )\, , \cr
\hat{D}_{\Delta_2 \Delta_1 \Delta_4 \Delta_3}(z, \bar z ) &= 
|z|^{\Delta_1-\Delta_2-\Delta_3+\Delta_4}\hat{D}_{\Delta_1 \Delta_2 \Delta_3 \Delta_4}({z}, {\bar{z}} )\, ,
 \end{align}
which also imply, for instance, 
  \begin{align} \label{eq:dhat3}
\hat{D}_{\Delta_1 \Delta_3 \Delta_2 \Delta_4} \left({z\over z-1}, {\bar{z} \over \bar{z}-1} \right) &= 
|1-z|^{2\Delta_4}\hat{D}_{\Delta_1 \Delta_2 \Delta_3 \Delta_4}({z}, {\bar{z}} )\,, \cr
\hat{D}_{\Delta_3 \Delta_1 \Delta_2 \Delta_4} \left({z-1\over z}, {\bar{z}-1 \over \bar{z}} \right) &= 
|z|^{2\Delta_4}\hat{D}_{\Delta_1 \Delta_2 \Delta_3 \Delta_4}({z}, {\bar{z}} ) \, .
 \end{align}

\section{The Mellin amplitudes} \label{app:Mellin}
Following the convention of ref. \cite{Rastelli:2017udc}, the Mellin amplitude of the connected part of a correlation function is defined through
\begin{align} \label{eq:Mellin}
\mathcal{G}(U,V)=&{\pi  \over 2}  \int  {ds \over 4\pi i}{dt \over 4\pi i} \,U^{s\over 2} V^{{t - \Delta_{23} \over 2} } \mathcal{M}(s,t)\, \Gamma\left({\Delta_{12}-s \over 2}\right)\Gamma\left({\Delta_{34}-s \over 2}\right) \cr
& \times \, \Gamma\left({\Delta_{14}-t \over 2}\right) \Gamma\left({\Delta_{23}-t \over 2}\right) \Gamma\left({\Delta_{13}-u \over 2}\right)\Gamma\left({\Delta_{24}-u \over 2}\right),
\end{align} 
where the cross ratios $U, V$ are related to $z, \bar{z}$ via $U=|1-z|^2, V=|z|^2$, and $\mathcal{M}(s,t)$ is the Mellin amplitude. We have defined $\Delta_{ij}=\Delta_i + \Delta_j$. Here $s,t, u$ are variables in Mellin space, which satisfy the constraints $s+t+u=\sum_{i=1}^4 \Delta_i\equiv \hat\Delta$. In the flat space limit, they play the role of Mandelstam variables of scattering amplitudes. 

To compute the Mellin amplitudes of the two terms in (\ref{eq:Gcont}) in the main text, we will use the Mellin transformation of the $\hat{D}$-function, which is given by
  \begin{align} 
&\hat{D}_{\Delta_1 \Delta_2 \Delta_3  \Delta_4}(z,\bar{z}) = \Gamma\left({\hat{\Delta} -d \over 2}\right) {\pi^{d/2} \over 2 \prod_{j=1}^4 \Gamma(\Delta_j)} 
\int {ds \over 4\pi i} {dt \over 4\pi i} U^{s\over 2}  V^{t\over 2} \,\Gamma\left(-{s\over 2}\right) \Gamma\left(-{t\over 2}\right) \\
&\times \, \Gamma\left(\Delta_4 +{s+t \over 2}\right)\Gamma\left({\Delta_{12}  -\Delta_{34} -s \over 2}\right)  \Gamma \left({ \Delta_{23}  -\Delta_{14} -t \over 2} \right) \Gamma\left({\Delta_{134} - \Delta_{2} +s + t \over 2}\right)\,, \nonumber
 \end{align}
 where $\Delta_{ijk}=\Delta_i + \Delta_j + \Delta_k$. Starting with the term proportional to $\hat{D}_{1111}$, we have
  \begin{align} 
|1-z|^2 \hat{D}_{1111}(z,\bar{z}) &={\pi  \over 2} 
\int {ds \over 4\pi i} {dt \over 4\pi i} V^{t\over 2} U^{{s\over 2} +1}\,\Gamma^2\left(-{s\over 2}\right) \Gamma^2\left(-{t\over 2}\right) \Gamma^2\left(1 +{s+t \over 2}\right) .
 \end{align}
Shifting the integration variables by $t\rightarrow t-2$ and $s\rightarrow s-2$, we obtain
  \begin{align} 
|1-z|^2 \hat{D}_{1111}(z,\bar{z}) &={\pi  \over 2} 
\int {ds \over 4\pi i} {dt \over 4\pi i} V^{t -2\over 2} U^{{s\over 2}\, }\Gamma^2\left(-{s\over 2}+1\right) \Gamma^2\left(-{t\over 2}+1\right) \Gamma^2\left(-{u \over 2} +1\right) .
 \end{align}
Comparing with (\ref{eq:Mellin}), we see that the Mellin amplitude of $|1-z|^2 \hat{D}_{1111}(z,\bar{z})$ is simply~$1$. 
 
 Let us now consider the other contribution in (\ref{eq:Gcont}):
   \begin{align} 
|1-z|^2|z|^2 \hat{D}_{2112}(z,\bar{z}) &=  {\pi  \over 2  } 
\int {ds \over 4\pi i} {dt \over 4\pi i} V^{{t\over 2}+1} U^{{s\over 2}+1}\, \Gamma^2\left(-{s\over 2}\right) \Gamma\left(-{t\over 2}\right) \cr
&\times \, \Gamma^2\left(2 +{s+t \over 2}\right) \Gamma \left( -{t \over 2}-1 \right)\,.
 \end{align}
A similar change of integration variables ($t\rightarrow t-4$ and $s\rightarrow s-2$) leads to 
    \begin{align} 
|1-z|^2|z|^2 \hat{D}_{2112}(z,\bar{z}) =&  {\pi  \over 2  } 
\int {ds \over 4\pi i} {dt \over 4\pi i} V^{{t\over 2}-1} U^{{s\over 2}} \,\Gamma^2\left(-{s\over 2}+1\right) \Gamma^2\left(-{t\over 2}+1\right) \cr
&\times \, \Gamma^2\left(-{u \over 2}+1\right) \times   \left(-{t\over 2}+1\right)\,,
 \end{align}
so the Mellin amplitude is $-{t\over 2}+1$. Therefore, we conclude that the contributions given in (\ref{eq:Gcont}), when expressed in terms of Mellin space, lead to a linear combination of contact terms with at most two-derivatives and with $s\leftrightarrow u$ symmetry: $t$ and a constant. 
 
\providecommand{\href}[2]{#2}\begingroup\raggedright\endgroup

\end{document}